\documentclass[
aps,
twocolumn,
prb,
longbibliography,
groupedaddress,
superscriptaddress,
letterpaper,
amsfonts,
floatfix
]
{revtex4-1}
\RequirePackage{amsmath,amssymb,bm,wasysym}
\RequirePackage{graphicx}
\RequirePackage{hyperref}
\RequirePackage{braket}
\hypersetup{
breaklinks={true},
citecolor={blue},
colorlinks={true},
linkcolor={blue},
pdfcreator={\LaTeX and\flqq hyperref\frqq},
pdffitwindow={true},
pdfmenubar={true},
pdfpagelayout={OneColumn},
pdfstartview={FitH},
pdftoolbar={true},
plainpages={false},
pdfpagemode={UseThumbs},
}
\usepackage[usenames]{color}
\usepackage{booktabs}
\usepackage[table,xcdraw]{xcolor}
\usepackage{multirow}
\usepackage{rotating,tabularx}
\usepackage[table]{xcolor}
\DeclareMathAlphabet{\mathpzc}{OT1}{pzc}{m}{it}
\renewcommand{\vec}[1]{\boldsymbol{#1}}

\newcommand{\bea}{\begin{eqnarray*}}
\newcommand{\eea}{\end{eqnarray*}}
\newcommand{\bne}{\begin{equation*}}
\newcommand{\ede}{\end{equation*}}

\newcommand{\bnen}{\begin{equation}}
\newcommand{\eden}{\end{equation}}
\newcommand{\bean}{\begin{eqnarray}}
\newcommand{\eean}{\end{eqnarray}}
\newcommand{\bsen}{\begin{subequations}}
\newcommand{\esen}{\end{subequations}}

\newcommand{\bna}{\begin{array}}
\newcommand{\eda}{\end{array}}
\newcommand{\bnm}{\begin{enumerate}}
\newcommand{\edm}{\end{enumerate}}
\newcommand{\bni}{\begin{itemize}}
\newcommand{\edi}{\end{itemize}}

\renewcommand{\vec}[1]{\text{\boldmath{$ #1 $}}}

\begin{document}
\pagestyle{plain}
\title{Hyperfine-assisted decoherence of a 
phosphorus nuclear-spin qubit in silicon}
\author{Bence Het\'enyi}
\email{bence.hetenyi@unibas.ch}
\affiliation{Institute of Physics, E\"{o}tv\"{o}s University, 1518 Budapest, Hungary}
\affiliation{Department of Physics, University of Basel, Klingelbergstrasse 82, CH-4056 Basel, Switzerland}
\author{P\'eter Boross}
\affiliation{Institute for Solid State Physics and Optics, Wigner Research Centre for Physics, Hungarian Academy of Sciences, H-1525 Budapest P.O. Box 49, Hungary}
\author{Andr\'{a}s P\'{a}lyi}
\email{palyi@mail.bme.hu}
\affiliation{Department of Theoretical Physics
and MTA-BME Exotic Quantum Phases Research Group,
Budapest University of Technology and Economics, 1111 Budapest, Hungary}
\pacs{
}

\newcommand{\noteandras}[1]{\textcolor{red}{#1}}
\newcommand{\notebence}[1]{\textcolor{blue}{#1}}

\begin{abstract}
The nuclear spin of a phosphorus atom in silicon 
has been used as a quantum bit in various
quantum-information experiments. 
It has been proposed that this nuclear-spin
qubit can be efficiently controlled by an ac electric field,
when embedded in a two-electron dot-donor setup subject to 
intrinsic or artificial spin-orbit interaction. 
Exposing the qubit to control electric fields in that setup exposes
it to electric noise as well. 
In this work, we describe the effect of electric noise
mechanisms, such as phonons and $1/f$ charge noise, 
and estimate the corresponding decoherence
time scales of the nuclear-spin qubit. 
We identify a promising 
parameter range where the electrical single-qubit
operations are at least an order of magnitude faster then the decoherence.
In this regime, decoherence 
is dominated by dephasing due to $1/f$ charge noise.  
Our results facilitate the optimized design of nanostructures
to demonstrate electrically driven nuclear spin resonance.
\end{abstract}
\maketitle

\section{Introduction}

The nuclear spin of a phosphorus  (P) atom in silicon (Si)
is a highly coherent two-level system\cite{Kane,Zwanenburg,Steger,Muhonen,Saeedi},
and has been used as a qubit in 
several quantum-information experiments
\cite{Pla_nuc,FreerQMemory,DehollainBell,MuhonenWeak,Laucht}. 
Single-qubit control of such a nuclear spin 
have been demonstrated using ac magnetic fields, in the spirit of 
nuclear magnetic resonance\cite{Pla_nuc};
initialization and readout can be performed using the 
donor electron spin\cite{Morello,Pla_electron,Pla_nuc}. 

A recent work\cite{boross2018hyperfine} 
proposes an engineered nanostructure,
in which the P nuclear-spin qubit could be efficiently 
controlled by an ac electric field, offering several practical
advantages\cite{boross2018hyperfine,Tosi_natcomm,Sigillito,Thiele,Godfrin,tosi2018robust},
in comparison with the traditionally used ac magnetic field.
The proposed setup is a dot-donor structure
with two bound electrons\cite{Lansbergen,Urdampilleta,HarveyCollard,Rudolph},
and the interaction between the 
control electric field and the nuclear spin is mediated by
the interplay of hyperfine interaction and (intrinsic or artificial)
spin-orbit interaction.
In short, the ac electric field makes the spin of the electrons
time dependent, and hyperfine interaction translates the
time-dependent electronic spin to a time-dependent Knight field,
felt by the nuclear spin due to the hyperfine interaction. 
Hence, the system of the two electrons functions as a transducer,
converting the ac electric field to an effective ac magnetic field
for the nucleus. 

The above mechanism is useful for electrical control, but it also
exposes the nuclear-spin qubit to decoherence due to
electric fluctuations. 
In this work, we describe two electrical noise mechanisms, 
phonons and $1/f$ charge noise, 
in the setup described above.
Our goal is to estimate the corresponding
decoherence time scales of the nuclear-spin qubit. 
We identify a promising 
parameter range where the electrical single-qubit
operations are at least an order of magnitude faster than the decoherence, 
and the latter is dominated by dephasing due to 
$1/f$ charge noise. 
Our study complements earlier theory works where 
the decoherence of electron-spin and flip-flop qubits 
in the dot-donor system were described\cite{Tosi_natcomm,boross2016valley,huang2018spin}.

The rest of the paper is organized as follows. 
In Section \ref{sec:oneoverfdephasing}, we introduce the model of the 
dot-donor setup, and describe dephasing of the nuclear-spin
qubit due to $1/f$ charge noise, which is the dominant
information-loss mechanism in the considered range of 
parameters. 
Even though having two bound electrons in the dot-donor
system has definite advantages over having only a single electron, 
we also discuss the latter case, 
for completeness and because of its conceptual 
simplicity. 
In Section \ref{sec:chargenoiserelax},
 we analyse qubit relaxation processes and 
leakage from the qubit subspace due to $1/f$ charge noise. 
In Section \ref{sec:phononrelax},
 we describe how phonons contribute 
to relaxation and leakage.
Estimates of time scales 
for the various setups and mechanisms are collected
in Table \ref{tab: t1t2}.
A few remarks are made in Section \ref{sec:discussion},
and conclusions are drawn in Section \ref{sec:conclusions}.

\section{Nuclear-spin dephasing due to $1/f$ charge noise}
\label{sec:oneoverfdephasing}

In this section, we introduce the P:Si nuclear-spin qubit
and the special dot-donor setup that enables its electrical 
control.
We discuss two different arrangements:\cite{boross2018hyperfine}. 
\emph{Single-electron setup (1e)}:
A single electron 
is confined in the dot-donor system. 
This brings the advantage
of conceptual simplicity, but the quality of electric control 
suffers strongly from electrical noise.
\emph{Two-electron setup (2e)}: 
Two electrons are confined in the system, 
allowing for efficient electrical control even in the presence
of realistic charge noise. 
Our main goal then is to describe the dephasing of the nuclear-spin
qubit due to $1/f$ charge noise in both setups, to calculate
the dephasing time $T_2^*$, and to identify a parameter range 
where the dephasing time is longer than the time scale
of single-qubit operations.
This is indeed possible for the 2e setup, as revealed by the
last line of Table \ref{tab: t1t2}: 
the estimated dephasing rate for that particular working point
is $\Gamma^*_2 \equiv 1/T_2^* = 2.97$ kHz, whereas 
the Rabi frequency characterizing the operation time scale
is $f_\text{Rabi} \approx  53$ kHz.

\begin{figure*}
\centering
\hspace{-0.06\columnwidth}
\includegraphics[width=1.8\columnwidth]{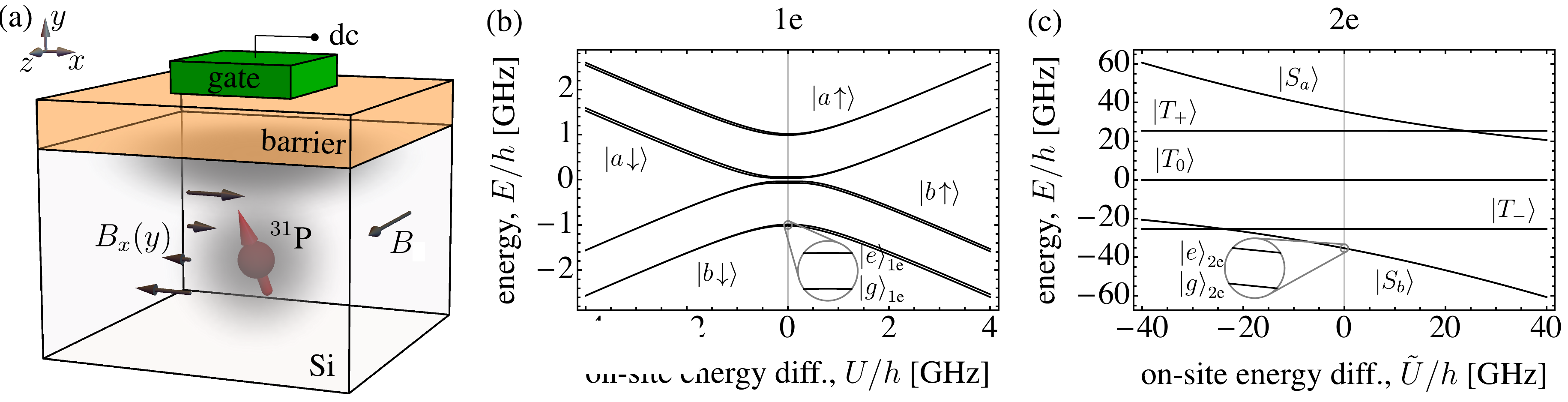}
\caption{
{\bf Dot-donor setup and the energy spectra of the 
coupled electron-nuclear system.}
(a) Dot-donor setup. 
The dc gate voltage is used to balance the donor electron (or electrons)
on a bonding orbital (gray cloud) of the artificial molecule
formed by the dot-donor system. 
Magnetic field (denoted by black arrows) has a homogeneous 
part $B$ along the $z$ direction and an inhomogeneous $B_x(y)$ along 
$x$. 
The inhomogeneous magnetic field can be 
replaced by a sufficiently strong spin-orbit interaction. 
Red arrow represents the nuclear spin.
(b) Energy spectrum of the coupled electron-nuclear system 
in the single-electron setup. 
Homogeneous magnetic field: $B = 35.7\, \text{mT}$,
tunnelling amplitude: $V_t/h = 1\, \text{GHz}$. 
(c) Energy spectrum in the two-electron setup. 
Magnetic field:
$B = 906.5\, \text{mT}$ , 
tunneling amplitude: 
$V_t/h = 50\, \text{GHz}$.
Zoom-ins in (b) and (c) show the two basis states of the
nuclear-spin qubit.}
\label{fig: mod_levs}
\end{figure*}

\subsection{Nuclear-spin qubit with a single donor electron}

The setup is shown in Fig.~\ref{fig: mod_levs}.
A P atom (red sphere) is embedded in a Si crystal, at distance 
$d$ from the interface with an insulating barrier (e.g. SiO$_2$).
The barrier separates the bulk Si from the gate electrode, 
the latter being used to control the position of the donor electron.
An ac component of the voltage on the gate electrode can
be used to electrically drive the nuclear-spin qubit; note, however,
that in this work we consider only the non-driven case, when 
the gate voltage is dc.

At low temperature and zero gate voltage, there is a single electron
bound to the donor nucleus. 
A finite dc gate voltage creates an electric field $\vec E(\vec r)$ 
in its vicinity.
When this electric field is pulling the electron toward the gate
strongly enough, 
then the electron is removed from the donor and sticks to the 
interface with the barrier, where it is trapped in a quantum-dot-like
confinement potential created by the gate electrode. 
Under certain conditions\cite{Tosi_natcomm}, 
there is a finite gate-voltage value which
ensures that half of the electronic wave function is localized on the
donor, and the other half is at the interface; this situation
is depicted in Fig.~\ref{fig: mod_levs}a, where the gray 
cloud corresponds to this `split' wave function resembling 
a bonding state in a diatomic molecule. 
We refer to this 
setting as the \emph{ionization point}\cite{Tosi_natcomm}. 

If the gate voltage is tuned to the vicinity of the ionization point, 
then the `orbital' or `charge'  
degree of freedom of the electron can be 
described using the two localized orbitals 
$\ket{i}$ and $\ket{d}$;
the former is the one localized at the interface, 
the latter is the one localized on the donor. 
Then, the  $2\times 2$ effective Hamiltonian that describes the 
charge in the vicinity of the ionization point can 
be written as\cite{Tosi_natcomm}
\begin{equation}
H_\text{ch} = \frac{U}{2}\sigma_z + \frac{V_t}{2}\sigma_x \, ,
\end{equation}
where $U$ is the on-site energy difference between the 
interface and donor orbitals (controlled by the gate voltage), 
$V_t$ is the tunnelling amplitude, and the Pauli matrices are defined in 
the basis above, e.g. as 
$\sigma_z = |i\rangle \langle i| - |d\rangle \langle d|$. 
We denote the eigenstates of $H_\text{ch}$ 
as $|a\rangle$ and $|b\rangle$, 
as a reference to 
the anti-bonding (higher-energy) and the 
bonding (lower-energy) state.
Note that a low-energy excited orbital, i.e., 
the valley pair of $\ket{i}$, is
available at the 
interface\cite{Tosi_natcomm,HarveyCollard,huang2018spin}, with an excitation energy varying
from a few tens to a few hundreds of microelectronvolts.
We disregard this state in our minimal model, assuming that
its excitation energy is much larger than the tunnel coupling.

The two-dimensional minimal model for the electron 
charge, introduced above, has to be extended with the electronic
spin 
and nuclear spin degrees of freedom, yielding an $8\times 8$
Hamiltonian
\begin{equation}
H = H_\text{ch} + H_{B,\text{e}} + H_{B,\text{n}}
 + H_\text{hf} + H_{\mu,\text{e}} + H_{\mu,\text{n}}.
 \label{eq: ham}
\end{equation}
This incorporates the effects of the
homogeneous static magnetic field
($H_{B}$),
hyperfine interaction  
between the electronic and nuclear spin ($H_\text{hf} $), 
and intrinsic or artificial spin-orbit interaction
($ H_{\mu}$).
In this work, we consider artificial spin-orbit interaction 
and neglect the intrinsic mechanism (see Section \ref{sec:discussion}
for a discussion).
More precisely, we assume the presence of 
an inhomogeneous magnetic field
along x, $B_x = \beta y$  (see Fig.~\ref{fig: mod_levs}a),
where the origin of the $y$ axis is chosen halfway between
the charge centers of the $\ket{i}$ and $\ket{d}$ orbitals. 
Then, the Hamiltonian terms read
\begin{subequations}
\begin{eqnarray}
H_{B,\text{e}} &=& h \gamma_\text{e} B
S_z
\\
H_{B,\text{n}} &=& -h \gamma_\text{n} B I_z \\
H_\text{hf} &=& A n_\text{d} \vec S \cdot \vec I
\\
H_{\mu,\text{e}} &=& h \gamma_\text{e} \frac{\beta d}{2} \sigma_z
S_x
\\
H_{\mu,\text{n}} &=& h \gamma_\text{n} \frac{\beta d}{2} I_x \, ,
\end{eqnarray}
\end{subequations}
where $A/h = 117\, \text{MHz}$ is the hyperfine coupling strength, 
while $\gamma_\text{e} = 27.97 \, \text{GHz}/\text{T}$ and 
$\gamma_\text{n} = 17.23 \, \text{MHz}/\text{T}$ are the electron  
and nuclear gyromagnetic factors. 
Furthermore, $\vec B = (0,0,B)$ is the homogeneous magnetic field,
the operator $n_\text{d} = (1-\sigma_z)/2$ is the occupation number of 
the donor site, 
$\vec S=(S_x,S_y,S_z)$ is the spin of the donor electron, and 
$\vec I = (I_x,I_y,I_z)$ is the 
nuclear spin of the P atom. 
We will denote the eigenstates of $S_z$ ($I_z$) 
as $\ket{\uparrow}$ and $\ket{\downarrow}$ 
($\ket{\Uparrow}$ and $\ket{\Downarrow}$).

Below, we will obtain analytical results for various physical quantities 
using perturbation theory.
In those calculations, the unperturbed Hamiltonian 
is chosen as
$H_0 = H_\text{ch} + H_{B,\text{e}} + H_{B,\text{n}}
 + H_\text{hf,sec}$, 
 whereas 
 the perturbation is $H_1 = H - H_0$. 
Above, $H_\text{hf,sec} = A n_\text{d} S_z I_z$ 
is the `secular' or `diagonal' part of 
the hyperfine interaction.
We call the eigenstates of $H_0$  \emph{unperturbed states}.
These are product states formed from the previously defined 
charge and spin states, 
and therefore we denote them as, e.g., 
$\ket{b\! \downarrow \Uparrow}_0$. 
As long as the perturbation matrix elements between the
unperturbed states are weak,
 the eigenstates of $H$ can be labelled with the same 
labels, such as $\ket{b\! \downarrow \Uparrow}$.

For a particular parameter set studied in 
Ref.~\onlinecite{boross2018hyperfine}, 
the energy eigenstates of $H$ are shown as the
function of the on-site energy difference $U$ in 
Fig.~\ref{fig: mod_levs}b. 
The basis states of the nuclear-spin qubit
are those with the two lowest energies
$\epsilon_g$ and $\epsilon_e$, 
highlighted by the zoom-in 
in Fig.~\ref{fig: mod_levs}b, denoted as 
$|g \rangle_{1e} \equiv \ket{b\! \downarrow \Uparrow}$ and 
$|e\rangle_{1e} \equiv \ket{b\! \downarrow \Downarrow}$.

\begin{table*}
\resizebox{1.\textwidth}{!}{
\begin{tabular}{|c|c|c|c|c|c|c|c|c|c|c|c|c|c|}
\cline{2-10}
 \multicolumn{1}{c|}{\multirow{2}{*}{}}& Dephasing [Hz] & \multicolumn{4}{c|}{Relaxation [Hz]} & \multicolumn{4}{c|}{Leakage [Hz]} & \multicolumn{4}{c}{\multirow{2}{*}{}} \\ 
\cline{2-14}
\multicolumn{1}{c|}{} &\rule{0pt}{9pt} $\Gamma_2^*$ & $\Gamma^{e}_\text{R,p}$ & $\Gamma^{g}_\text{R,p}$ & $\Gamma^{e}_\text{R,c}$ & $\Gamma^{g}_\text{R,c}$ & $\Gamma^{e}_\text{L,p}$ & $\Gamma^{g}_\text{L,p}$ & $\Gamma^{e}_\text{L,c}$ & $\Gamma^{g}_\text{L,c}$ & $B\, \text{[T]}$ & $V_t / h$ & $f_\text{Rabi}$& $f_\text{L}$ \\ 
\hline 
\rule{0pt}{9pt} 1e  & \cellcolor{red!25}$4.52 \times 10^7$ & $4.79 \times 10^{-6}$ & $4.65 \times 10^{-6}$ & $1110$ & $1070$ & $1380$ & $1500$ & \cellcolor{red!25}$9.05 \times 10^{6}$ & \cellcolor{red!25}$1.09 \times 10^{6}$ & $0.0357$ & $1\, \text{GHz}$ & $72\, \text{kHz}$&$31.2\, \text{MHz}$\\ 
\hline 
\rule{0pt}{9pt} 2e  & \cellcolor{blue!25}$2970$ & $2.35 \times 10^{-10}$ & $2.31 \times 10^{-10}$ & $0.437$ & $0.43$ & $6.97 \times 10^{-3}$ & $5.46 \times 10^{-3}$ & $1.01 \times 10^{-2}$ & $8.07 \times 10^{-3}$ & $0.907$ & $50\,  \text{GHz}$ & $53\, \text{kHz}$ & $15.6\, \text{MHz}$\\ 
\hline 
\end{tabular}
}
\caption{
Qubit types, working points, and information-loss time scales due
to different mechanisms in the P:Si dot-donor system. 
$\Gamma_\text{R}$: relaxation rate, 
$\Gamma_\text{L}$: leakage rate,
$\Gamma_2^*$ dephasing rate.
Lower index "p" ("c") refers to phonons ($1/f$ charge noise). 
For relaxation and leakage processes, the
initial state can be the qubit ground state $| g \rangle$ or 
the qubit excited state $| e \rangle$. 
Rates are evaluated for temperature $T = 50\, \text{mK}$.}
\label{tab: t1t2}
\end{table*}

\subsection{Nuclear-spin dephasing due to 1/f charge noise}

Our goal here is to focus on the nuclear-spin qubit, 
and describe the information-loss mechanisms 
arising from its interaction with electrical fluctuations. 
Recent experiments\cite{Freeman,Yoneda}
on state-of-the-art silicon quantum devices
have shown pronounced significance of 
$1/f$ charge noise.
By comparing various information-loss mechanisms, we will conclude 
that the most relevant one
is  dephasing  due to $1/f$ charge noise.

To describe dephasing, in this section, we treat the $1/f$ charge noise 
as a time-dependent on-site energy difference $\delta U(t)$ 
between the interface and the donor felt by the electron. 
Correspondingly, the noise Hamiltonian reads
\bean
\label{eq:noisehamiltonian1e}
H_\text{noise} = \frac{\delta U(t)}{2} \sigma_z.
\eean
We describe the dynamics of the nuclear-spin qubit under the 
influence of this noise using the effective qubit Hamiltonian,
obtained by projecting the total Hamiltonian
$H_\text{tot} = H + H_\text{noise}$ on the 
qubit subspace:
\begin{eqnarray}
\begin{split}
 H_\text{q}=P H_\text{tot} P = 
 \frac{h f_\text{L}}{2} \sigma_z'
+ \frac{\delta U(t)}{2}
\mathcal{L}
\sigma_z' + 
\frac{\delta U(t)}{2}
\mathcal{T}
\sigma_x'
\end{split}
\label{eq: qubham}
\end{eqnarray}
where $P = |e\rangle \langle e| + |g\rangle \langle g|$, and
the qubit Larmor frequency $f_\text{L}$, the longitudinal 
matrix element $\mathcal{L}$, 
and the transverse matrix element $\mathcal{T}$, 
are defined via
\begin{subequations}
\label{eq:params1e}
\bean
h f_\text{L} &=& \epsilon_e-\epsilon_g, \\
\label{eq:longitudinal1e}
\mathcal{L} &=& 
\frac{\langle e | \sigma_z |e \rangle - \langle g | \sigma_z |g \rangle}{2}, \\
\mathcal{T} &=& \langle e | \sigma_z |g \rangle.
\eean
\end{subequations}
Furthermore, the Pauli matrices are defined in the two-dimensional
qubit subspace, e.g., $\sigma_z' = \ket{e}\bra{e} - \ket{g}\bra{g}$.
In general, both the longitudinal $\mathcal{L}$ 
and transverse $\mathcal{T}$ matrix elements are finite. 
However, for describing dephasing, 
it is a good approximation to disregard $\mathcal{T}$
in our setup, and keep only $\mathcal{L}$; this is what we 
do from now on. 
The validity of this approximation is discussed in 
section \ref{sec:discussion}. 

We assume that the noise is 
classical, Gaussian\cite{makhlin2003dissipation,Kogan}
and has the $1/f$ power spectrum\cite{Petit,Freeman,Jung_apl,Dekker}
\begin{equation}
\label{eq:alphadef}
S_{\delta U} (f) \equiv
\int_{-\infty}^{\infty} d t \, e^{i2\pi f t} \, \overline{\delta U(t) \delta U(0)}
=
\frac{\alpha_{1/f} k_B T}{2\pi f} \, ,
\end{equation}
where the overline denotes the average over the noise
realizations, $\alpha_{1/f}$ denotes the overall strength
of the noise, $k_B$ is the Boltzmann constant, and $T$ is the
temperature. 
From the analysis of Refs.~\onlinecite{Tosi_natcomm}
(Ref.~\onlinecite{Yoneda}), we estimate 
$\alpha_{1/f} \approx 43.5 \,  \text{neV}$ 
($\alpha_{1/f} \approx 5.1 \, \text{neV}$), 
see Appendix \ref{app:noiseestimates}.
In what follows, we will take the greater value for $\alpha_{1/f}$
to obtain quantitative results. 

In the description of dephasing due to $1/f$ noise, the experimental
integration time $t_i$ is often 
taken into account\cite{makhlin2003dissipation}. 
For example, the decay of the qubit polarization in a Ramsey-type
experiment\cite{Pla_nuc} is inferred by doing many measurement
cycles: there are many grid points along the waiting-time axis, 
and for each waiting time, many measurement cycles are carried
out to obtain reliable statistics. 
The integration time $t_i$ is the total time required to carry out all these 
measurement cycles; in our estimates, we will assume that
the order of magnitude of this integration time is a second. 
The finite integration time implies that the slow noise components,
i.e., those that can be regarded constant
during the integration time,  
will not be able to influence the experiment. 
This argument implies that dephasing is insensitive to 
the low-frequency part of the noise spectrum, that is,
the part of the  spectrum below
the integration frequency $f_i = 1/t_i$ can be
neglected. 

With the above assumptions, dephasing can be
characterized by an approximately Gaussian 
decay\cite{makhlin2003dissipation}, 
i.e., the length of the qubit polarization vector 
decays in time as $\sim e^{-\left(\Gamma_2^* t\right)^2}$,
with the inhomogeneous dephasing rate given by
\begin{equation}
\Gamma_2^* \approx \frac{E_{1/f}}{h}\sqrt{2\pi\ln \frac{E_{1/f}}{hf_i}}\, ,
\label{eq: makhlindeph}
\end{equation}
where 
we have introduced 
$E_{1/f} = \sqrt{\alpha_{1/f}k_BT\, } \mathcal{L}$.
The energy scale
$E_{1/f}$ can be expressed from the
above numerical estimate of the noise strength $\alpha_{1/f}$; 
assuming $T = 50$ mK, we
find $E_{1/f} \approx \, h  \times
\mathcal{L} \times  0.1\, \text{GHz} $.
Note that a necessary formal condition for
the approximations leading to Eq.~\eqref{eq: makhlindeph}
is  $\frac{E_{1/f}}{h f_i} \gtrsim 10$.
In practice, this condition always holds (as long as this is the dominant dephasing mechanism), since it is translated by
Eq.~\eqref{eq: makhlindeph} to 
$\Gamma_2^* / f_i \gtrsim 40$, and the latter holds because in 
any reasonable dephasing-time measurement, 
the integration time is orders of magnitude larger than 
the dephasing time itself.

We evaluate the result Eq.~\eqref{eq: makhlindeph} 
for a particular parameter set (see Table \ref{tab: t1t2})
using the nuclear-spin basis states 
obtained by the numerical diagonalization of 
$H$ of Eq.~\eqref{eq: ham} and the subsequent evaluation of 
$\mathcal{L}$ from Eq.~\eqref{eq:longitudinal1e}.
Our numerical estimate for this particular working point 
is $\Gamma_2^* = 45.2 \, \text{MHz}$, a rate that
is much larger than the esimated electrically induced Rabi 
frequency $f_\text{Rabi} = 72$ kHz at this working point.
(For the Rabi frequency estimate, see
Ref.~\onlinecite{boross2018hyperfine}.)
This is a strong indication that the nuclear spin cannot
be used as a qubit in this working point.
One way to make this a useful qubit is to reduce
the strength of $1/f$ noise by at least 4 orders of magnitude.

The reason for this strong dephasing is as follows. 
Recall that the working point studied here is the ionization point. 
Here, a weak electrical perturbation can displace the
electron along the dot-donor (y) direction very effectively,
which implies a significant change in the electron density
on the donor, which in turn implies a significant
change in the Knight shift felt by the nuclear spin.
This mechanism was implicitly quantified already in
Eq.~(5) of Ref.~\onlinecite{boross2018hyperfine}, 
see the third component of $\vec{b}_\text{ac}$ therein.

\subsection{Nuclear-spin qubit and its dephasing with two donor electrons}

In contrast to the 1e setup showing poor coherence properties,
much improvement is anticipated for the
2e setup\cite{boross2018hyperfine}. 
Importantly, 
silicon-based dot-donor devices holding an even number of electrons
are available experimentally\cite{HarveyCollard,harvey2018high}.
Here we introduce the model Hamiltonian for this system
following Ref.~\onlinecite{boross2018hyperfine},
recall how electrical control of the nuclear-spin qubit 
is envisioned, 
and determine
the dephasing rate of the nuclear-spin qubit 
due to $1/f$ charge noise. 

The two electrons in the dot-donor system can fill 
the orbitals $\ket{i}$ and $\ket{d}$, both providing 
two sublevels due to the electron spin.
This implies that there are six two-electron states to
take into account in a minimal model.
We use the standard basis set
$\ket{S_{20}}$, 
$\ket{S}$, 
$\ket{T_+}$,
$\ket{T_0}$,
$\ket{T_-}$,
$\ket{S_{02}}$.
Here, the first [last] element is the spin singlet state
in which both electrons are localized at the interface
[donor], also referred to as the (2,0) [(0,2)] charge configuration.
Furthermore, the remaining four states are the standard singlet and
triplet states in which one electron is localized at each site,
also referred to as the (1,1) charge configuration. 

We assume that the on-site Coulomb repulsion $U_C$
between the electrons ($\sim$ meV)
is much larger then the tunnel coupling $V_t$
($\sim 10 - 100\, \mu$eV).
In the 2e setup, we consider the case when  
the on-site energy difference $U$ is set
such that on-site energies of the four (1,1) states
and the single (0,2) state
are close to each other, 
and we restrict our attention to the dynamics in this 
low-energy
electronic subspace.
That is, we neglect the high-energy $\ket{S_{20}}$ 
state, 
and model the system by a ten-dimensional Hamiltonian, 
using the five electronic and the two nuclear-spin
basis states.

The Hamiltonian describing this arrangement is 
the two-electron version of the single-electron Hamiltonian introduced in 
Eq.~\eqref{eq: ham}.
The terms which are different from the single-electron case read:
\begin{subequations}
\begin{eqnarray}
\label{eq: 2echargeham}
H_\text{ch} &=& - \tilde{U} \ket{S_{02}}\bra{S_{02}} + \frac{V_t}{\sqrt{2}} \left( \ket{S}\bra{S_{02}} + h.c. \right),
\\
H_{B,\text{e}} &=& h \gamma_\text{e} B \left( \ket{T_+}\bra{T_+} - \ket{T_-}\bra{T_-} \right) ,
\\
H_{\mu,\text{e}} &=& h \gamma_\text{e} \frac{\beta d}{2 \sqrt{2}} \left( \ket{T_-}\bra{S} - \ket{T_+}\bra{S} \right) + h.c. ,
\label{eq:inhom2e}
\end{eqnarray}
\end{subequations}
where $\tilde{U} = U-U_C$ is the energy detuning measured from the $(1,1)-(0,2)$ tipping point $U=U_C$.
Each terms above acts as the identity on the nuclear spin. 
Furthermore, the hyperfine Hamiltonian takes the form
\begin{eqnarray}
\begin{split}
H_\text{hf} =&\quad{} \frac{A}{2} ( \ket{T_+}\bra{T_+} -\ket{T_-}\bra{T_-}\\
 &\quad{} - \ket{S}\bra{T_0} - \ket{T_0}\bra{S})  I_z \\
 &\quad{} + \frac{A}{2\sqrt{2}}  [(\ket{S}\bra{T_+} + \ket{T_0}\bra{T_+} \\
 &\quad{} + \ket{T_-}\bra{T_0} - \ket{T_-}\bra{S})  I_+ + h.c].
\end{split}
\label{eq:2ehyperfine}
\end{eqnarray}
Note that in Eq.~\eqref{eq:2ehyperfine}, we have corrected a few typos
that appeared in Eq.~(A2b) of Ref.~\onlinecite{boross2018hyperfine}.

Following the 1e case, it is useful to introduce the 
anti-bonding and bonding singlet energy eigenstates of $H_\text{ch}$,
i.e., the molecular states formed by $\ket{S}$ and $\ket{S_{02}}$, 
which we will denote as $\ket{S_a}$ and $\ket{S_b}$, respectively.

We plot the ten energy eigenvalues
of this model in Fig~\ref{fig: mod_levs}c, 
as a function of the detuning parameter 
$\tilde{U}$, for homogeneous magnetic field 
$B = 906.5\, \text{mT}$, tunnelling amplitude 
$V_t/h = 50\, \text{GHz}$,
magnetic-field gradient $\beta = 0.47 \, \text{mT}/\text{nm}$,
and donor-interface distance $d = 15 \, \text{nm} $.
Here again, we define the nuclear-spin qubit basis states as 
the lowest-energy eigenstates:
$|e\rangle_{2e} \equiv \ket{S_b\! \Downarrow}$ and $|g\rangle_{2e} \equiv \ket{S_b\! \Uparrow}$. 

Before presenting our results for the dephasing time caused
by $1/f$ charge noise, we recall that an important energy 
scale
for the nuclear-spin qubit is the 
energy gap between the electronic 
states $\ket{S_b}$ and $\ket{T_-}$, see Fig.~\ref{fig: mod_levs}c. 
We denote the value of this gap at zero detuning
$\tilde{U} = 0$, 
as obtained from the electronic Hamiltonian
$H_\text{ch} + H_{B,\text{e}}$,
by $\delta$. 
It can be expressed as
\bean
\label{eq:delta}
\delta = \frac{V_t}{\sqrt{2}} - h \gamma_\text{e} B.
\eean
To ensure that the nuclear-spin qubit dynamics upon 
electrical drive follows regular and fast Rabi 
oscillations\cite{boross2018hyperfine}, 
it is reasonable to set the value of $\delta$ much
larger than the coupling matrix element
induced by the inhomogeneous magnetic 
field between the electronic states $\ket{S_b}$ and 
$\ket{T_-}$.
This relation is satisfied, e.g., with the choice
$\delta = 200 \braket{S_b | H_{\mu, \text{e}} | T_-}$.
The parameter values given above satisfy this relation,
since they correspond to 
$\braket{S_b | H_{\mu, \text{e}} | T_-} \approx 50\, \text{MHz}$ 
and $\delta \approx 10 \, \text{GHz}$.

In the 2e setup, the noise Hamiltonian, derived from 
its 1e counterpart in Eq.~\eqref{eq:noisehamiltonian1e},
reads
\bean
H_\text{noise} = - \delta U(t) \ket{S_{02}}\bra{S_{02}}.
\eean
Projecting the total Hamiltonian onto 
the two-dimensional subspace of the nuclear-spin qubit,
as done for the 1e case in Eq.~\eqref{eq: qubham}, 
we obtain the right hand side of Eq.~\eqref{eq: qubham}
with the following identifications:
\begin{subequations}
\label{eq:params2e}
\bean
h f_\text{L} &=& \epsilon_e-\epsilon_g, \\
\label{eq:longitudinal2e}
\mathcal{L} &=& 
\left| \braket{g | S_{02} } \right|^2
- \left|\braket{e | S_{02}}\right|^2
,\\
\mathcal{T} &=& - 2
\braket{e| S_{02}} \braket{S_{02} | g}.
\eean
\end{subequations}

We obtain the dephasing rate $\Gamma_2^*$ 
from Eq.~\eqref{eq: makhlindeph}, 
after evaluating the longitudinal matrix
element $\mathcal{L}$ in Eq.~\eqref{eq:longitudinal2e}
with the numerically obtained energy eigenstates
$\ket{g}$ and $\ket{e}$.
For the above parameter values, 
we find $\Gamma_2^* \approx 2.97\, \text{kHz}$, 
see Table \ref{tab: t1t2}. 
Note that this rate is significantly smaller than the Rabi frequency
in Table \ref{tab: t1t2} (estimated in Ref.~\onlinecite{boross2018hyperfine}),
suggesting that the nuclear spin can be used as a functional qubit in 
this setting.

Now, we argue that this dephasing 
is a consequence of the hyperfine interaction, 
and not influenced significantly by the inhomogeneous 
magnetic field. 
This is revealed by a perturbative approach 
that yields the analytical result
for the longitudinal coupling strength 
\begin{eqnarray}
\begin{split}
\label{eq:Lpert}
\mathcal{L} =& \frac{A^2}{32 \delta^2} +\frac{2 A^2}{32 \left(2\Delta_Z + 2\delta \right)\delta}\\
&-\frac{2 A^2}{32 \left(2\Delta_Z + \delta \right)\left(2 \Delta_Z + 2 \delta \right)}-\frac{A^2}{32 \left(2\Delta_Z + \delta \right)^2},
\end{split}
\end{eqnarray}
with $\Delta_Z = h \gamma_\text{e} B$ and correspondingly, 
$\Gamma_2^* \approx 2.77 \, \text{kHz}$
in the working point defined above. 
The result \eqref{eq:Lpert} depends on the hyperfine coupling strength $A$,
but does not depend on the magnetic-field gradient $\beta$. 

To obtain Eq.~\eqref{eq:Lpert}, 
we take 
$H_0 = H_\text{ch} + H_{B,\text{e}} + H_{B,\text{n}}$ as 
the unperturbed Hamiltonian, 
 take $H_1 = H-H_0 - H_{\mu,\text{n}}$ 
as the perturbation, and
neglect the small term $H_{\mu,\text{n}}$ for simplicity. 
Then, we apply time-independent third-order perturbation theory
to calculate the perturbation-induced change in the energy 
splitting of the nuclear-spin qubit states
$\ket{S_b \Downarrow}$ and
$\ket{S_b \Uparrow}$.
According to Eq.~\eqref{eq: qubham}, 
we identify this change with $\delta U \, \mathcal L $,
express $\mathcal{L}$, and use $h \gamma_\text{n} B \ll h \gamma_\text{e} B, V_t, \delta$ to obtain Eq.~\eqref{eq:Lpert}.

The perturbative result \eqref{eq:Lpert} for dephasing mechanism 
is visualized in the 
level diagram shown in Fig.~\ref{fig: 2edephmatrixelements}.
The blue horizontal lines depict the energy levels of the unperturbed
Hamiltonian. 
Arrows represent relevant perturbation matrix elements that contribute to 
dephasing. 
In the third-order formula \eqref{eq:Lpert}, each of the four 
terms can be associated to 
a three-step loop drawn by the perturbation matrix elements.
For example, the loop corresponding to the first term is
$\ket{S_b \Downarrow}_0 \rightarrow \ket{S_b \Downarrow}_0
\rightarrow \ket{T_- \Uparrow}_0 \rightarrow \ket{S_b \Downarrow}$.

To extend the perturbative formula \eqref{eq:Lpert} of $\mathcal{L}$
for the case of non-zero $\tilde U$, we use
the exact eigenvalues and eigenstates of \eqref{eq: 2echargeham}, 
yielding
\begin{eqnarray}
\begin{split}
\label{eq: 2edeph}
\mathcal{L} =& \frac{2V_t^2}{\tilde{U}^2 + 2V_t^2} \left[ \frac{A^2}{32 \delta'^2} +\frac{2 A^2}{32 \left(2\Delta_Z + 2\delta' \right)\delta'} \right.\\
& \left.-\frac{2 A^2}{32 \left(2\Delta_Z + \delta' \right)\left(2\Delta_Z + 2\delta' \right)}-\frac{A^2}{32 \left(2\Delta_Z + \delta' \right)^2}\right],
\end{split}
\end{eqnarray}
where 
$\delta ' = \frac{\tilde{U}}{2} + \frac{\sqrt{\tilde{U}^2 + 2V_t^2}}{2} - \Delta_Z$.
In comparison to Eq.~\eqref{eq:Lpert}, the energy denominators
are different in \eqref{eq: 2edeph}.
The prefactor in Eq.~\eqref{eq: 2edeph} depends on the squared
detuning parameter $\tilde U^2$, i.e., this prefactor is
suppressed if we move away from 
the working point from $\tilde U = 0$ either to the positive or negative
direction.
For the negative [positive] direction, 
this dephasing suppression is dominated 
by the feature that the electronic state acquires a growing weight
in the (1,1) [(0,2)] charge configuration, 
and hence gets less affected by noise (hyperfine interaction).

\begin{figure*}
\centering
\hspace{-0.06\columnwidth}
\includegraphics[width=2\columnwidth]{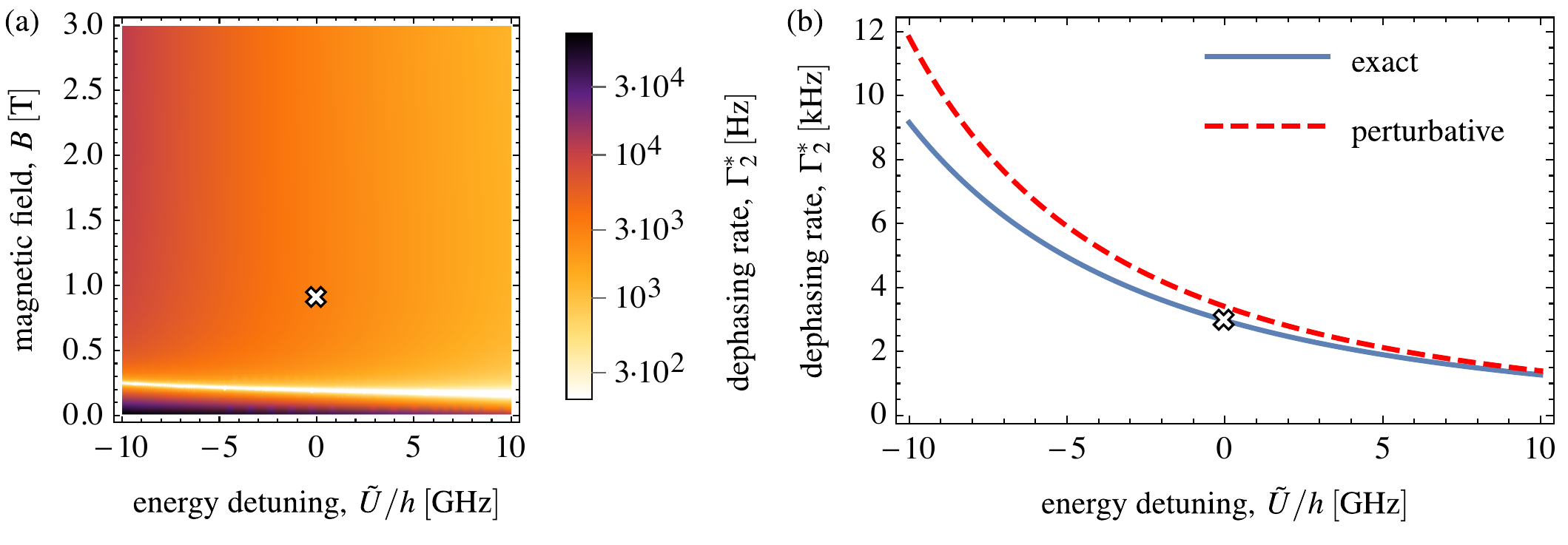}
\caption{
Dephasing rate induced by $1/f$ charge noise in 
the two-electron setup.  (a) Numerically calculated
dephasing rate 
as a function of the (1,1)-(0,2) energy detuning
and the magnetic field. 
Note that the tunnel matrix element $V_t$ is tied
to the magnetic field $B$ to ensure a constant
$S$-$T_-$ energy gap at $\tilde U = 0$, via
$V_t =\sqrt{2} \left( \delta + h \gamma_\text{e} B \right)$ 
(cf. Eq.~\eqref{eq:delta}).
The white cross denotes the working point 
$B = 906.5\, \text{mT}$, 
$\tilde{U}=0$ and 
$\delta/h = 10\, \text{GHz}$. 
(b) 
Exact numerical result (solid) is compared to the 
perturbative result 
of Eq.~\eqref{eq: 2edeph}
at the working-point magnetic field $B = 906.5\, \text{mT}$.
Further parameters: $\alpha_{1/f} = 43.5\, \text{neV}$, 
$T = 50\, \text{mK}$.}
\label{fig: deph}
\end{figure*}

\begin{figure}
\includegraphics[width=\columnwidth]{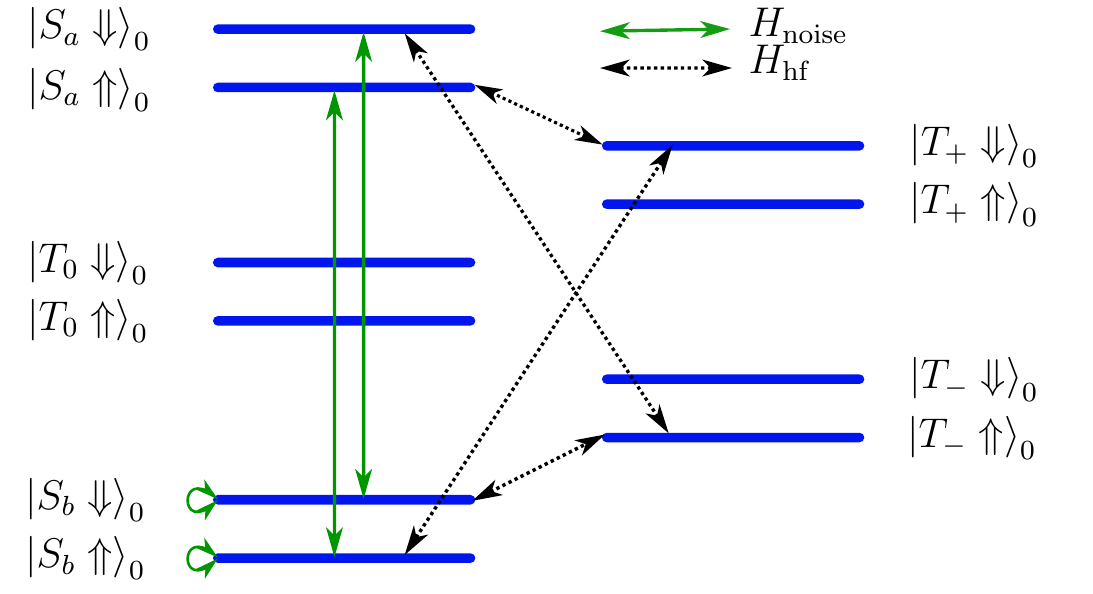}
\caption{Level diagram of the hyperfine states in 
the two-electron setup.
Horizontal lines indicate the eigenstates of the
unperturbed Hamiltonian $H_\text{ch} + H_{B,\text{e}}+ H_{B,\text{n}}$.
Arrows indicate the perturbation matrix elements 
relevant for the dephasing in the specific working point (see text).
Perturbation matrix elements that are not relevant for this dephasing
mechanism are not shown.} \label{fig: 2edephmatrixelements}
\end{figure}

In Fig.~\ref{fig: deph}a, we plot the numerically calculated
dephasing rate $\Gamma_2^*$ as a function of
the energy detuning $\tilde U$ and the magnetic 
field $B$. 
In the figure, the above working point is denoted by 
an `x'. 
Note that in the figure,
the tunnel matrix element $V_t$ is changed
together with the magnetic field $B$ such that
$\delta$ is kept fixed, see Eq.~\eqref{eq:delta}.
The most relevant features
in Fig.~\ref{fig: deph}a are as follows.

(i) For $B\gtrsim0.4$ T, the dephasing rate decreases as the energy
detuning $\tilde U$ increases, and the dephasing rate is
hardly dependent on the magnetic field.  
(ii) These trends are confirmed by the perturbative result \eqref{eq: 2edeph}. 
For example, a comparison of the numerical and perturbative 
result, 
along a horizontal cut of Fig.~\ref{fig: deph}a 
containing the working point `x', 
is shown in Fig.~\ref{fig: deph}b.
Using Eq.~\eqref{eq: 2edeph}, we can explain that the decreasing
trend of the dephasing rate with increasing energy detuning $\tilde U$ 
is mostly due to the increasing energy gap between the electronic
singlet ground state and the electronic excited states, 
cf.~Fig.~\ref{fig: mod_levs}c.
(iii) For magnetic fields much weaker than the working point
value, the dephasing rate does depend significantly
on $B$.  
This regime is beyond the validity of the perturbative
result \eqref{eq: 2edeph} due the smallness of the
magnetic field.

In conclusion, we have evaluated the inhomogeneous dephasing
rate of the nuclear-spin qubit for a P:Si 1e and 2e dot-donor setup
subject to (artificial) spin-orbit coupling,
and identified a parameter range for the 2e setup where the
dephasing time is much longer than the time required for single-qubit
operations. 

\section{Relaxation and leakage due to 1/f charge noise}
\label{sec:chargenoiserelax}

\begin{figure}
\includegraphics[width=0.7 \columnwidth]{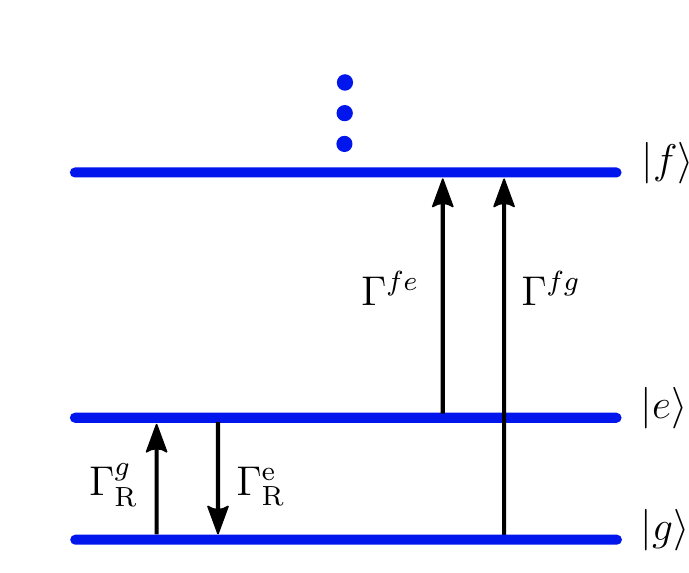}
\caption{
Inelastic transitions due to $1/f$ charge noise
and phonons.
Downhill and uphill processes within
the qubit subspace contribute to qubit relaxation. 
Transitions from one of the
qubit basis states $\ket{g}$ or $\ket{e}$ to
a state outside of the qubit subspace cause
leakage.}\label{fig:relaxationandleakage}
\end{figure}

Besides dephasing discussed above, 
the presence of $1/f$ charge noise also opens up
channels for information loss. 
Here, we focus on the 2e setup, 
and describe two types of inelastic processes
caused by $1/f$ charge noise, 
see Fig.~\ref{fig:relaxationandleakage}.
First, we consider inelastic processes between the 
two qubit basis states, 
denoted as $\Gamma^{e}_\text{R}  \equiv \Gamma^{ge} $ and 
$\Gamma^{g}_\text{R}  \equiv \Gamma^{eg} $ in Fig.~\ref{fig:relaxationandleakage},
to be referred to as \emph{relaxation}.
Second, we consider inelastic processes 
that bring the system from one of the qubit basis states
to a state outside the qubit's Hilbert space, 
to be referred to as \emph{leakage}, shown
as $\Gamma^{fg}$ and $\Gamma^{fe}$
in Fig.~\ref{fig:relaxationandleakage}.
Our conclusion is that the time scales of these
processes in the vicinity of the working point 
are much longer than the dephasing time
$1/ \Gamma_2^*$ derived in the previous section,
hence they hardly affect the functionality of the nuclear-spin qubit. 

We use the qubit Hamiltonian $H_\text{q}$ of 
Eq.~\eqref{eq: qubham} to describe 
the relaxation processes, with the adjustment that 
$\delta U$ is treated now as an operator representing 
the environment producing the $1/f$ noise. 
Recall that 
the longitudinal and transverse coupling matrix elements 
$\mathcal{L}$ and $\mathcal{T}$ 
are evaluated for the 1e [2e] setup
via Eq.~\eqref{eq:params1e} 
[Eq.~\eqref{eq:params2e}].
According to Bloch-Redfield
theory\cite{makhlin2003dissipation}, 
the downhill and uphill relaxation rates 
are given by
\begin{subequations}
\begin{eqnarray}
\Gamma^{e}_\text{R} &=& \frac{1}{2\hbar^2} 
\mathcal{T}^2 
\frac{1+n_\text{BE}(h f_\text{L})}{1+2n_\text{BE}(h f_\text{L})} \, 
S_{\delta U} (f_\text{L})\, ,
\\
\Gamma^{g}_\text{R} &=& \frac{1}{2\hbar^2} 
\mathcal{T}^2\, \frac{n_\text{BE}(h f_\text{L})}{1+2n_\text{BE}(h f_\text{L})}\, S_{\delta U} (f_\text{L}).
\label{eq:uphillrelaxation}
\end{eqnarray}
\end{subequations}
Here, $n_\text{BE}(h f_\text{L})$ is the temperature-dependent 
Bose-Einstein function, and
$S_{\delta U}(f)$ is the symmetrized noise power spectrum of
the operator $\delta U$, which is given by the same formula
\eqref{eq:alphadef}
as in the classical case.

For the 2e setup, the numerical values of the 
$1/f$-induced relaxation rates of the nuclear-spin qubit
are around 1 Hz, see Table \ref{tab: t1t2}. 
This implies that relaxation due to $1/f$ noise 
is much less relevant than dephasing. 
Note that the uphill and downhill relaxation rates are almost the
same, in line with the fact that the thermal frequency 
scale $k_B T/h \approx 1.04\,  \text{GHz}$ well exceeds
the 
qubit splitting $f_\text{L} \approx 14.5\, \text{MHz}$ 
in this point. 

The transverse matrix element can be obtained 
from perturbation theory similarly as in the case of dephasing.

One possible way for the derivation 
is to apply quasiquasidegenerate perturbation theory\cite{Winkler}
to obtain an effective Hamiltonian for the nuclear-spin qubit subspace,
identify the off-diagonal element of that Hamiltonian with 
$\delta U\,  \mathcal{T}/2$ according to Eq.~\eqref{eq: qubham},
and express $\mathcal{T}$ from that equation.
From this approach, for the tipping point $\tilde U = 0$,
we obtain
\bean
\label{eq:Tpert}
\mathcal{T} &=& 
\frac{A (h \gamma_e \beta d)}{16} 
\left( \frac{1}{\delta^2} + \frac{1}{(2\Delta_Z+\delta )^2} \right.
\nonumber
\\
&+& \left. 
\frac{2}{(2\Delta_Z+2\delta)\delta} +\frac{2}{(2\Delta_Z+2\delta)(2\Delta_Z+\delta)} \right).
\eean
The value obtained from this formula is 
$\mathcal{T} \approx 1.96 \times 10^{-5}$,
in good agreement with the numerical result
$\mathcal{T} = 1.98 \times 10^{-5}$.

\begin{figure}
\includegraphics[width=\columnwidth]{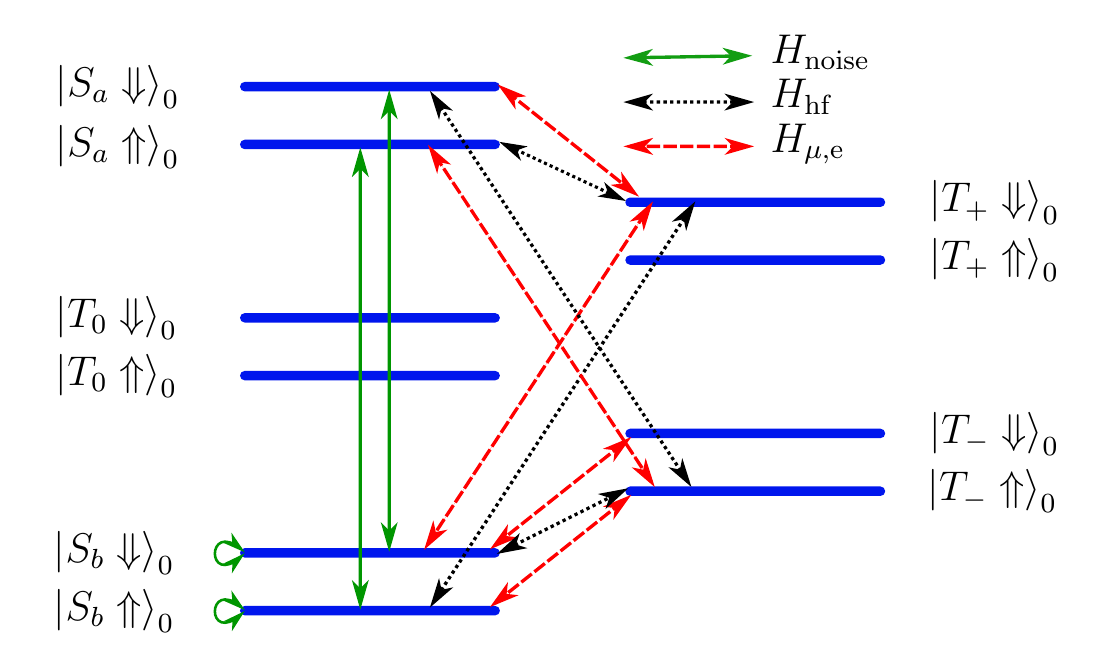}
\caption{Level diagram of the hyperfine states in 
the two-electron setup.
Horizontal lines indicate the eigenstates of the
unperturbed Hamiltonian $H_\text{ch} + H_{B,\text{e}}+ H_{B,\text{n}}$.
Arrows indicate the perturbation matrix elements involved in relaxation and leakage (see text). }\label{fig: 2erelleakmatrixelements}
\end{figure}

The perturbative contributions can again be visualized
by drawing the perturbation matrix elements as steps between
the energy levels of the unperturbed Hamiltonian; this is shown in
Fig.~\ref{fig: 2erelleakmatrixelements}.
The four terms in the perturbative formula \eqref{eq:Tpert}
correspond to 
six three-step paths in Fig.~\ref{fig: 2erelleakmatrixelements}
connecting the two qubit basis states.
For example, the first term of \eqref{eq:Tpert}
corresponds to the path 
$\ket{S_b \Downarrow}_0 \to \ket{S_b \Downarrow}_0 \to 
\ket{T_- \Uparrow}_0 \to \ket{S_b \Uparrow}_0$.
As seen in Fig.~\ref{fig: 2erelleakmatrixelements}, all
three-step paths connecting the two qubit basis states contain
one hyperfine matrix element and one inhomogeneous magnetic field
matrix element (besides one noise matrix element), hence
we conclude that relaxation in this case is dominated by the
interplay of hyperfine interaction and the inhomogeneous magnetic field.

Leakage rates, i.e., noise-induced transition 
rates from the nuclear-spin qubit basis states toward
higher-lying eigenstates
(see Fig.~\ref{fig:relaxationandleakage}), can also be described
by the Bloch-Redfield result. 
For example, the leakage from the qubit ground
state $\ket{g}$ has the rate
\begin{eqnarray}
\label{eq:leakageg}
\Gamma_\text{L}^{g} &=& 
\sum_{f\neq g,e} \Gamma^{fg}
\\ \nonumber
&=&\frac{1}{2\hbar^2} \sum \limits_{f\neq g,e} 
\mathcal{T}_{fg}^2\,  \frac{n_\text{BE}(\epsilon_{fg})}{1+2n_\text{BE}(\epsilon_{fg})} \, S_{\delta U} (\epsilon_{fg}/h).
\end{eqnarray}
Here, the sum goes for the possible higher-lying
final states $\ket{f}$, 
the matrix element is
$\mathcal{T}_{fg} = \braket{f | \sigma_z |g }$ for
the 1e case and 
$\mathcal{T}_{fg} = -2 \braket{f | S_{02}}\braket{S_{02} |g }$
for the 2e case, and 
$\epsilon_{fg}$ is the distance between the energies
of $\ket{f}$ and $\ket{g}$. 
The leakage rate $\Gamma_L^{e}$ for the qubit
excited state $\ket{e}$ is expressed analogously 
to Eq.~\eqref{eq:leakageg}.

As seen in Table \ref{tab: t1t2}, the leakage
rates for the 2e setup in the working point
are of the order of $0.01$ Hz, slower and hence
less significant than the 
previously considered processes.
At this working point, the dominant leakage
process from $\ket{g}$ is the one
toward $\ket{T_- \Uparrow}$.
Approximating the leakage-rate sum in
Eq.~\eqref{eq:leakageg} with this single 
contribution, expressing 
$\mathcal{T}_{fg}$ using 
perturbation theory, and 
using the relation $\delta \ll h \gamma_\text{e} B$, 
we find
\bean
\Gamma_\text{L}^g \approx
 \frac{1}{2\hbar^2} \frac{(h\gamma_e \beta d)^2 }{8\delta^2}
 \frac{1}{e^{(\delta /k_B T)}+1} \frac{\alpha_{1/f} k_B T}{\delta}.
\eean
According to this formula, leakage is dominated by 
the magnetic-field gradient $\beta$. 
Similar considerations lead to $\Gamma_L^e \approx
\Gamma_L^g \left[1+( \frac{A}{h\gamma_e \beta d})^2 \right]$, 
i.e., that the leakage from the excited state has an additional
contribution from the hyperfine interaction $A$.
These results are 
in line with the level diagram shown in 
Fig.~\ref{fig: 2erelleakmatrixelements},
where the qubit ground state is hybridized with a single
$T_-$ sublevel, 
whereas the qubit excited state is hybridized with both
$T_-$ sublevels.

\section{Relaxation and leakage due to phonons}
\label{sec:phononrelax}

Besides charge noise,
phonon absorption or emission can also cause transitions between the energy levels. 
Here we consider the deformation-potential 
electron-phonon interaction mechanism\cite{Herring,Yu_Cardona}. 
This mechanism is enhanced in a silicon dot-donor
electron system 
(compared to, e.g., a double-dot or double-donor setup)
due to the different valley compositions
of the electronic states in the dot and the donor\cite{boross2016valley}.
For the energy distances considered here, only 
long-wavelength acoustic phonons have to be 
considered. 

We describe the phonon-mediated inelastic transitions
using Bloch-Redfield theory, similarly to the case
of charge-noise-mediated transitions in section
\ref{sec:chargenoiserelax}. 
As a natural consequence, the phonon-induced
transition rates are related to the charge-noise-induced
transitions rates by the relative weight of the 
noise spectral densities at the transition frequency. 
For example, the phonon-mediated downhill
relaxation rate is
\bean
\label{eq:inelasticphonon}
\Gamma^{e}_\text{R,p} =
\frac{S_\text{ph}(f_L)}{S_{\delta U}(f_L)}
\Gamma^{e}_\text{R,c},
\eean
and analogous relations hold for the phonon-mediated
uphill relaxation rate $\Gamma^{g}_\text{R,p}$ 
and the leakage rates 
$\Gamma^{e}_\text{L,p}$ and $\Gamma_\text{L,p}^{g}$,
with the caveat that $f_L$ in Eq.~\eqref{eq:inelasticphonon}
has to be replaced with the corresponding transition frequencies. 

To obtain the  noise spectral density 
$S_\text{ph}(f_L)$, 
representing the phonons, 
one starts from the 
single-electron 
electron-phonon interaction Hamiltonian\cite{boross2016valley}
for the dot-donor system:
\bean
H_\text{eph} = \frac{\hat{U}_\text{ph}}{2} \sigma_z,
\eean
where $\hat{U}_\text{ph}$ is expressed via phonon
creation and annihilation operators in 
Eqs. (12) and (13) of Ref.~\onlinecite{boross2016valley}.
The symmetrized phonon noise density is 
then expressed as (cf.~Eq.~\eqref{eq:alphadef})
\bean
\label{eq:quantumnoise}
S_\text{ph}(f) = \frac 1 2 
\int_{-\infty}^{\infty}
dt e^{i2\pi f t} 
	\overline{
	\left\{
		\hat{U}_\text{ph}(t), \hat{U}_\text{ph}(0)
	\right\}
	},
\eean
where the time-dependent operators are 
defined in the interaction picture, the curly brackets 
$\{.,.\}$ denote the anticommutator, 
and the overline denotes thermal average for
the equilibrium phonon bath\cite{Clerk}.

Equation \eqref{eq:quantumnoise} is
evaluated as
\begin{subequations}
\bean
S_\text{ph}(f) &=& 
\left[ 1+ 2n_\text{BE}(h f) \right] S_\text{ph}^{(0)}(f),\\
S_\text{ph}^{(0)}(f) &=&
\frac{\Xi^2 \hbar}{30 \pi \rho}
\left(
	\frac{2}{3v_\text{L}^5} + \frac{1}{v_\text{T}^5}
\right)
(2\pi f)^3.
\eean
\label{eq:phononnoise}
\end{subequations}
Here, the material-specific parameters for 
silicon are the uniaxial deformation potential parameter
$\Xi_u = 8.77\, \text{eV}$, 
the mass density 
$\rho = 2330 \, \text{kg}/\text{m}^3$, 
and the 
longitudinal and transverse sound velocities
$v_\text{L} = 9330 \text{m}/\text{s}$
and 
$v_\text{T} = 5420 \text{m}/\text{s}$.

The results obtained for the $1/f$ charge noise
model can therefore be converted to 
the case of phonon-mediated relaxation and 
leakage using Eqs.~\eqref{eq:phononnoise} and
\eqref{eq:inelasticphonon}.
As seen in Table \ref{tab: t1t2}, the phonon-induced
relaxation rates are approximately 10 orders of 
magnitude smaller then the relaxation rates
due to $1/f$ charge noise. 
In contrast, the leakage rates corresponding
to processes induced by phonons and $1/f$ charge noise
are very similar. 
We emphasize that all of these leakage and relaxation 
rates are smaller than the dephasing rate.

\section{Discussion}
\label{sec:discussion}

\emph{Artifical vs intrinsic spin-orbit interaction.}
Ref.~\onlinecite{boross2016valley} suggests that
electrical control of the nuclear-spin qubit should be possible
either by relying on an inhomogeneous magnetic field
(artificial spin-orbit interaction), or by relying on 
intrinsic spin-orbit interaction. 
In a simple phenomenological picture, 
spin-orbit interaction can influence
the dot-donor system in two ways; 
both effects have been observed
in silicon double quantum dots\cite{Veldhorst_spinorbit,jock2017probing,Tanttu,HarveyCollardSO}.
On the one hand, 
it  renormalizes the g-factor (with few percents), potentially
making it anisotropic and different at the donor
and in the dot.
On the other hand, it induces a spin-dependent
interdot tunnelling matrix element (few tens of MHz). 
The consequences of the anisotropic and
different g-factors are similar to those of the
inhomogeneous magnetic field. 
The consequences of the spin-dependent tunneling
term are expected to be qualitatively different. 
For example, while the inhomogeneous magnetic
field provides matrix elements within the 
(1,1) charge configuration, e.g., between
$\ket{S}$ and $\ket{T_-}$ (see Eq.~\eqref{eq:inhom2e}),
spin-dependent tunnelling provides a matrix element
connecting (1,1) states with $\ket{S_{02}}$.
Nevertheless, in the vicinity of the (1,1)-(0,2) tipping point, 
where the singlet electronic ground state
$\ket{S_g}$ is a balanced superposition of 
(1,1) and (0,2) charge states, we expect that the 
dynamics induced by spin-orbit interaction is similar to that
induced by an inhomogeneous magnetic field.

\emph{Neglecting the transverse noise term in the dephasing model.}
We calculated the nuclear-spin qubit dephasing time 
due to $1/f$ charge noise in section \ref{sec:oneoverfdephasing}.
For the dephasing calculation, we disregarded 
the transverse noise term with the prefactor
$\mathcal{T}$. 
This simplification is justified as long as the 
noisy component of the Larmor frequency
is dominated by the longitudinal component 
proportional to $\mathcal{L}$; formally, that condition
reads $\mathcal{T}^2 \delta U/(2 h f_L \mathcal L) \ll 1$.  
For a rough estimate of the importance of $\mathcal T$
in dephasing, we take $\delta U = 1\, \mu$eV, yielding
$\mathcal{T}^2 \delta U/(2 h f_L \mathcal L) \approx 6.70 \times 10^{-4}$
for the working point of the 2e setup shown in Fig.~\ref{fig: deph}a,
and thereby suggesting that our result is accurate in 
the vicinity of the working point.
Note, however that for low magnetic field, far from 
the working point, the longitudinal matrix element $\mathcal L$ 
vanishes, see the horizontal white stripe for $B \approx 0.2\, \text{T}$ in 
Fig.~\ref{fig: deph}a.
In this region, the description of dephasing should be 
refined\cite{Shnirman}.

\emph{Leakage due to uphill charge transitions.}
For the 2e setup, we have described
information loss at a well-defined working point
specified in, e.g., Table \ref{tab: t1t2}.
Departing from this example working point might
provide optimized results for certain target quantities, 
e.g., the qubit quality factor $f_\text{Rabi}/\Gamma_2^*$.
We leave such optimization for future work, hopefully
aided by input from experiments. 
Nevertheless, we do emphasize one important feature
that arises upon decreasing the tunnel coupling $V_\text{t}$
with respect to the working point discussed above. 
Namely, spin-conserving uphill charge transitions from 
$\ket{S_b}$ to $\ket{S_a}$ can destabilize the qubit
if the energy gap between the two electronic states 
is too narrow.
The corresponding leakage rate can be calculated, e.g., 
from Eq.~\eqref{eq:uphillrelaxation} by using
$\mathcal{T} \approx 1$ and
$f_\text{L} \mapsto \sqrt{2}V_t/h$.
For example, at a relatively
low tunnel coupling value $V_t = h \times 1.8$ GHz, 
this leakage rate is $\approx 1 MHz$, and is dominantly 
induced by $1/f$ charge noise.

\section{Conclusions}
\label{sec:conclusions}

In conclusion, we consider information-loss mechanisms
for an electrically controllable phosphorus nuclear-spin qubit 
in a silicon nanostructure. 
We identify a parameter set (working point)
where the information-loss time scales are longer than
the estimated control time, 
suggesting that this setup is suited to demonstrate
coherent electrical control of a nuclear spin.
In this working point, the dominant
decoherence mechanism is dephasing due to 
$1/f$ charge noise. 
Our results are expected to facilitate the optimized design
of nanostructures for quantum information experiments
with nuclear-spin qubits. 

\acknowledgments
We thank W.~A.~Coish, L.~Cywinski, P.~Harvey-Collard,
V. Srinivasa
for useful discussions. 
This research was supported by the National Research Development and Innovation Office of Hungary within the Quantum Technology National Excellence Program  (Project No. 2017-1.2.1-NKP-2017-00001), 
and Grant
124723.
A.~P.~
was supported by the New National Excellence Program
of the Ministry of Human Capacities.

\appendix

\section{Estimate of the strength of the $1/f$ charge noise}
\label{app:noiseestimates}

In this appendix, we provide the details on how 
we estimated the strength $\alpha_{1/f}$ of
the $1/f$ charge noise, as defined in 
Eq.~\eqref{eq:alphadef}.

Our first estimate is based on 
Ref.~\onlinecite{Tosi_natcomm}, which 
provides a realistic characterization of 
the power spectrum of $1/f$ 
electric-field fluctuation in 
silicon-based nanostructures similar to the one
considered in the present work. 
Namely, at $T=100$ mK, they use the electric-field
noise spectrum
\bean
S_E(f) = \frac{\beta_E}{2\pi f},
\eean
with $\beta_E = \frac 1 6 10^4\, \text{V}^2/\text{m}^2$.
This is converted to on-site energy fluctuation via
$S_{\delta U} = e^2 d^2 S_E$, where 
$d$ is the distance between the charge center of the
interface-bound charge state $\ket{i}$ and
the charge center of the donor-bound charge state
$\ket{d}$. 
Using this result for $S_{\delta U}$, 
Eq.~\eqref{eq:alphadef} as a definition for $\alpha_{1/f}$,
and the parameter values $d=15\, \text{nm}$ and
$T = 100 \, \text{mK}$, we can express
$\alpha_{1/f} \approx 43.5 \, \text{neV}$, as quoted in 
the main text below Eq.~\eqref{eq:alphadef}.

Our second estimate is based
on the recent electron spin qubit 
experiment reported in Ref.~\onlinecite{Yoneda}.
Their Fig.~4b shows the
spectral density $S_{\delta f_L}$ of the fluctuations 
$\delta f_L$ of the
spin-qubit Larmor frequency $f_L$.
From that log-log plot, we can read off that the data
is well described by the relation
$y \approx 6.5 - x$,
where $y = \log_{10} S_{\delta f_L}$, and
$x = \log_{10} f$.
This is directly converted to 
\bean
S_{\delta f_L}(f)  
\approx \frac {3 \times 10^6 \text{s}^{-2}}{f} 
\approx \frac{2 \times 10^7 \text{s}^{-2}}{2\pi f}
\equiv \frac{\beta_{\delta f_L}}{2\pi f} .
\label{eq:betavalue}
\eean

Ref.~\onlinecite{Yoneda} uses a simple model that 
establishes a linear relation between the qubit Larmor frequency
fluctuation $\delta f_L$ and the quantum-dot on-site energy
fluctuation $\delta U$, $\delta U = \gamma \delta f_L$.
This implies that the on-site energy fluctuation is
characterized by the spectral density 
\bean
S_{\delta U}(f) = \gamma^2 S_{\delta f_L}(f), 
\eean
which, together with our Eqs.~\eqref{eq:alphadef}
and \eqref{eq:betavalue}, 
yields
\bean
\label{eq:alphayoneda}
\alpha_{1/f} = \frac{\gamma^2 \beta_{\delta f_L}}{k_B T}.
\eean
To obtain an estimate for our target quantity
$\alpha_{1/f}$, we need to express $\gamma$ first. 
To this end,  we use the model of 
Ref.~\onlinecite{Yoneda},
which is based on the physical picture that
the $1/f$ noise is caused by fluctuating charge traps, 
located at a typical distance $d_\text{ct}$ from the
center of the quantum dot. 
Within their model, the scaling factor $\gamma$ 
is given by
\bean
\label{eq:gamma}
\gamma = 
\frac{h m \omega_0^2 d_\text{ct}}{g \mu_B b_\text{long}}.
\eean
Here, $m\approx 0.2 m_\text{e}$ is the relevant
conduction-band effective mass in silicon, 
$\hbar \omega_0$ is the orbital level spacing of the quantum
dot,
$g \approx 2$ is the effective electronic g-factor, and
$b_\text{long}$ is the micromagnet-induced
gradient of the longitudinal magnetic field.
Using Eqs.~\eqref{eq:betavalue} and \eqref{eq:gamma}
in Eq.~\eqref{eq:alphayoneda}, 
and inserting the parameter values
$\hbar \omega_0 = 1 \, \text{meV}$,
$b_\text{long} = 0.2 \, \text{mT}/\text{nm}$,
$d_\text{ct} = 100\, \text{nm}$, 
and $T=100\, \text{mK}$, 
we obtain the value $\alpha_{1/f} \approx 5.1\, \text{neV}$,
as quoted in 
the main text below Eq.~\eqref{eq:alphadef}.

\bibliography{nuclearspindecoherence}

\end{document}